\documentclass[conference]{IEEEtran}
\usepackage{cite}
\usepackage{amsmath,amssymb,amsfonts}
\usepackage{graphicx}
\usepackage{booktabs}
\usepackage{textcomp}
\usepackage{xcolor}
\usepackage{placeins}

\begin{document}

\title{Simulation-Based Performance Evaluation of Sharded Blockchain Architectures}

\author{
\IEEEauthorblockN{Om Amit Gandhi}
\IEEEauthorblockA{College of Computing\\
Illinois Institute of Technology, Chicago, IL\\
ogandhi1@hawk.iit.edu}
\and
\IEEEauthorblockN{Ioan Raicu}
\IEEEauthorblockA{College of Computing\\
Illinois Institute of Technology, Chicago, IL\\
iraicu@iit.edu}
}

\maketitle

\begin{abstract}
Public blockchains continue to struggle with scalability because improving throughput is not as simple as increasing block size or reducing block interval. Larger blocks increase validation and transmission cost, while shorter intervals raise the likelihood of propagation delays, forks, and stale blocks. These limits motivate sharding, where transaction processing is divided across multiple parallel shard groups. In this work, we present a configurable SimPy-based discrete-event simulator for evaluating sharded blockchain architectures under controlled workload and network assumptions. The simulator models mining, verification, inter-shard coordination, block dissemination, measured throughput, average block time, and communication overhead. Our simulator achieves 1.6M TPS at 256 shards under a local datacenter-like setup and 0.6M TPS in a global WAN setup, showing strong throughput gains from parallel execution. However, the gains are not unbounded: beyond a certain number of shards, coordination traffic, synchronization, and network overhead begin to dominate, leading to diminishing returns.
\end{abstract}


\section{Introduction}
Public blockchains continue to struggle with scalability because improving throughput is not as simple as increasing block size or reducing block interval. Larger blocks increase validation and transmission cost across the network, while shorter block intervals raise the likelihood of propagation delays, forks, and stale blocks, so neither knob can be tuned in isolation without introducing new bottlenecks. Sharding has emerged as a leading alternative, partitioning transaction processing across parallel shard groups so that aggregate throughput can scale with the number of shards \cite{skidanov2019near,zamani2018rapidchain}. In principle, doubling the number of shards should roughly double throughput, since each shard processes an independent subset of transactions in parallel. In practice, however, shards cannot operate in complete isolation: they must periodically coordinate to maintain a single consistent chain, and this coordination introduces its own latency and message overhead. Real deployments cannot offer the controlled, reproducible conditions needed to isolate this scaling behavior or to identify the crossover point where coordination overhead begins to dominate, since live networks mix protocol effects with hardware variance, real user load, and uncontrolled network conditions.
 
To address this gap, we built a configurable, SimPy-based discrete-event simulator that models a sharded blockchain architecture end-to-end: transaction submission, shard-level mining and verification, inter-shard coordination via a metronome process, block assembly, and dissemination under configurable network conditions. The simulator allows shard count, block size, block time, and network profile to be swept independently while holding all other parameters fixed, isolating the effect of each on measured throughput, block time, and communication overhead. This paper describes the simulator design and reports throughput results across shard counts and three representative network environments, ranging from a single datacenter to a global wide-area deployment.
 
\section{Motivation}
Bitcoin achieves approximately 7 transactions per second (TPS) \cite{nakamoto2008bitcoin}, and Ethereum sustains roughly 15--30 TPS, far below the throughput demands of global financial infrastructure such as card payment networks, which routinely handle tens of thousands of TPS. Naive approaches to scaling, such as increasing block size or decreasing block interval, introduce their own costs: larger blocks raise validation and propagation overhead, since every node in the network must download, verify, and relay the full block before mining can continue, while shorter intervals increase the rate of forks and stale blocks as competing miners are more likely to find valid blocks before the previous one has fully propagated. These effects compound rather than cancel, so monolithic chains tend to plateau at a throughput ceiling that cannot be lifted by parameter tuning alone.
 
These limits motivate sharding~\cite{skidanov2019near}, where parallel shard groups divide the workload so that each shard handles only a fraction of the global transaction volume, allowing aggregate throughput to scale with the number of shards. However, sharding introduces a new source of overhead: shards must periodically exchange messages to agree on a global ordering of blocks, and as shard count grows, so does the volume of this coordination traffic. The crossover point at which coordination overhead begins to outweigh the gains from parallelism remains poorly understood without a controlled experimental setup, since it depends jointly on shard count, block size, and the latency and bandwidth characteristics of the underlying network. Understanding this crossover is central to determining how many shards a sharded blockchain can practically support before further sharding stops paying off.

\section{Related Work}
Sharding designs such as NEAR \cite{skidanov2019near} and RapidChain \cite{zamani2018rapidchain} propose protocol-level mechanisms for partitioning state and transaction validation across shards, including cross-shard transaction routing and committee reconfiguration. NEAR's Nightshade design in particular preserves a single logical chain while distributing transaction execution across shards: each shard produces a "chunk" of transactions every round, and these chunks are aggregated into a single block header by the chain's block producer, so that the protocol exposes one canonical chain to external observers despite parallel execution underneath \cite{skidanov2019near}. This aggregation step is precisely the kind of coordination point our simulator targets, since the time required to collect and merge chunks from all shards before a block can be finalized scales with shard count and network latency, and can become the limiting factor on throughput at high shard counts even though per-shard execution itself remains fast.
 
Because wide-area, multi-node experiments are costly and difficult to reproduce, simulation has become a standard tool for studying these tradeoffs. BlockSim is a discrete-event simulation framework \cite{alharby2019blocksim} that models blockchain systems at the level of nodes, network links, and consensus rounds, allowing researchers to configure block intervals, block sizes, network topology, and consensus parameters and observe resulting throughput and latency without deploying a live network. BlockSim's general-purpose design makes it well suited to studying single-chain consensus tradeoffs, but it does not natively model the shard-to-shard coordination structure that sharded architectures introduce, such as a metronome or leader-aggregation process that must wait on multiple parallel shard winners before a block can be committed.
 
While these works establish the theoretical scaling potential of sharding and provide general-purpose simulation infrastructure for single-chain systems, empirical evaluation of the throughput--coordination tradeoff specific to sharded designs under varying network conditions typically requires either large testnet deployments or analytical models that abstract away implementation-level overhead. Our simulator complements this body of work by modeling the shard-level mining, verification, and metronome-based coordination structure directly, providing a lightweight, reproducible environment for sweeping shard counts, block sizes, and block times under different network profiles and directly measuring the resulting coordination overhead.

\section{Simulator Design}

\begin{figure}[h]
\centering
\includegraphics[width=0.8233\linewidth]{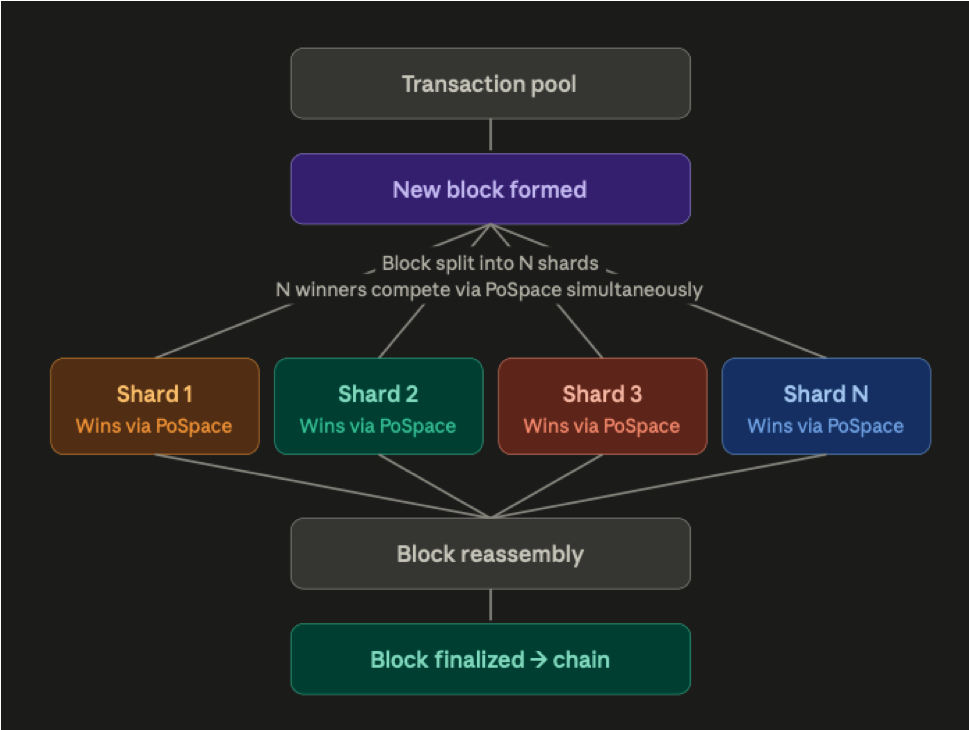}
\caption{Sharded architecture overview.}
\label{fig:arch}
\end{figure}

The simulator follows a MEMO-style architecture (Fig.~\ref{fig:arch}), implemented in Python 3.10 using SimPy 4.1.1 with shards, the metronome, and the transaction pool as independent processes communicating through events and shared stores. Wallets submit transactions to a shared pool; a metronome broadcasts per-round challenges and collects proof-of-space responses to coordinate timing, while shards compete in parallel, with each winner assembling a block from claimed transactions before block assembly merges per-shard results for submission to the chain. The design partitions execution across $S$ shards, each validating a disjoint subset of the pool under a shared target block time, with per-shard difficulty set via a harmonic-number correction so the slowest shard's expected time matches this target. Coordination uses a leader-based metronome: the first shard to finish broadcasts a $64$-byte announcement and becomes leader, while other winners send summaries only to it, which then aggregates results and triggers a global commit, decoupling per-shard validation from the commit and reducing blocking relative to epoch-based designs. Node and miner population scales with shard count ($100$ nodes up to $64$ shards, up to $1{,}000$ at $512$), with peers connected via random subsets and flooding, and the pool pre-filled to keep all configurations under sustained load.
 
Each round proceeds through mining, verification, coordination, and broadcast. Mining times are drawn from exponential distributions scaled by the harmonic correction; verification cost is fixed per transaction, with the slowest shard gating this phase. Coordination cost is modeled at the message and bandwidth level, with message counts derived analytically ($N$ announcements plus $\text{winners}-1$ summaries per round) and serialization delay from message size, count, bandwidth, and RTT for the chosen network profile (datacenter, US WAN, or global WAN). The leader's timing signal smooths block times while preserving parallelism, and since only the leader receives other winners' summaries, coordination traffic scales with shard count rather than transaction volume. Each transaction is assigned to exactly one shard, eliminating cross-shard transactions; double-spend prevention uses a two-phase scheme where shards validate locally and the leader applies a first-arrival conflict rule before an atomic commit.
 
Throughput is total committed transactions divided by simulated wall-clock time, and average block time is the mean inter-arrival time between committed blocks. Communication overhead is tracked as message volume and cumulative serialization/RTT delay relative to total simulation time, exposing the throughput--coordination crossover directly. Shard count, block size, block time, network condition, and verification cost are all configurable, with each configuration run as an independent simulation.

\FloatBarrier

\section{Experimental Setup}
All experiments were run on a machine with two Intel Xeon Silver 4108 CPUs (16 physical / 32 logical cores, up to 3.00 GHz turbo), 64 GB RAM, and 4$\times$ Samsung 970 EVO 250 GB NVMe SSDs, running Ubuntu 22.04.4 LTS (kernel 5.15.0-171) with Python 3.10.12 and SimPy 4.1.1.
 
Table~\ref{tab:params} summarizes the simulation parameters swept across experiments. Each configuration is defined by a (shard count, block size, mining time, network condition) tuple, and every combination is run as an independent simulation, yielding a full grid sweep across the parameter space rather than a single fixed configuration per network condition.
 
For each run, the simulator records measured TPS, average actual block time, and total coordination message volume after an initial warm-up period during which the transaction pool and shard states stabilize. TPS is computed from total committed transactions across all shards divided by simulated wall-clock time, and is compared against the configured mining time to quantify how far actual block time deviates from the target. Communication overhead is reported both as a raw message count and as a fraction of total simulation time spent in coordination-related delay, together exposing the point at which additional shards stop translating into additional throughput. Results are grouped by network condition and block size to isolate the effect of shard count on TPS, block time, and overhead while holding the remaining parameters fixed, with the same sweep repeated independently for each of the three network profiles to evaluate how coordination latency interacts with shard count under datacenter, US WAN, and global WAN conditions.
 
\begin{table}[h]
\caption{Simulation Parameters}
\label{tab:params}
\centering
\begin{tabular}{@{}ll@{}}
\toprule
\textbf{Parameter} & \textbf{Values} \\
\midrule
Shard count       & 1, 2, 4, 8, 16, 32, 64, 128, 256, 512 \\
Block size         & 1{,}024 -- 524{,}288 txns per block \\
Mining time         & 0.001s -- 1{,}200s \\
Network condition  & Local datacenter, US WAN, Global WAN \\
Architecture       & MEMO-style \\
Verification cost  & Configurable per shard \\
Bandwidth / RTT     & Configurable per network profile \\
\bottomrule
\end{tabular}
\end{table}

\section{Results}
\begin{figure}[!t]
\centering
\includegraphics[width=\linewidth]{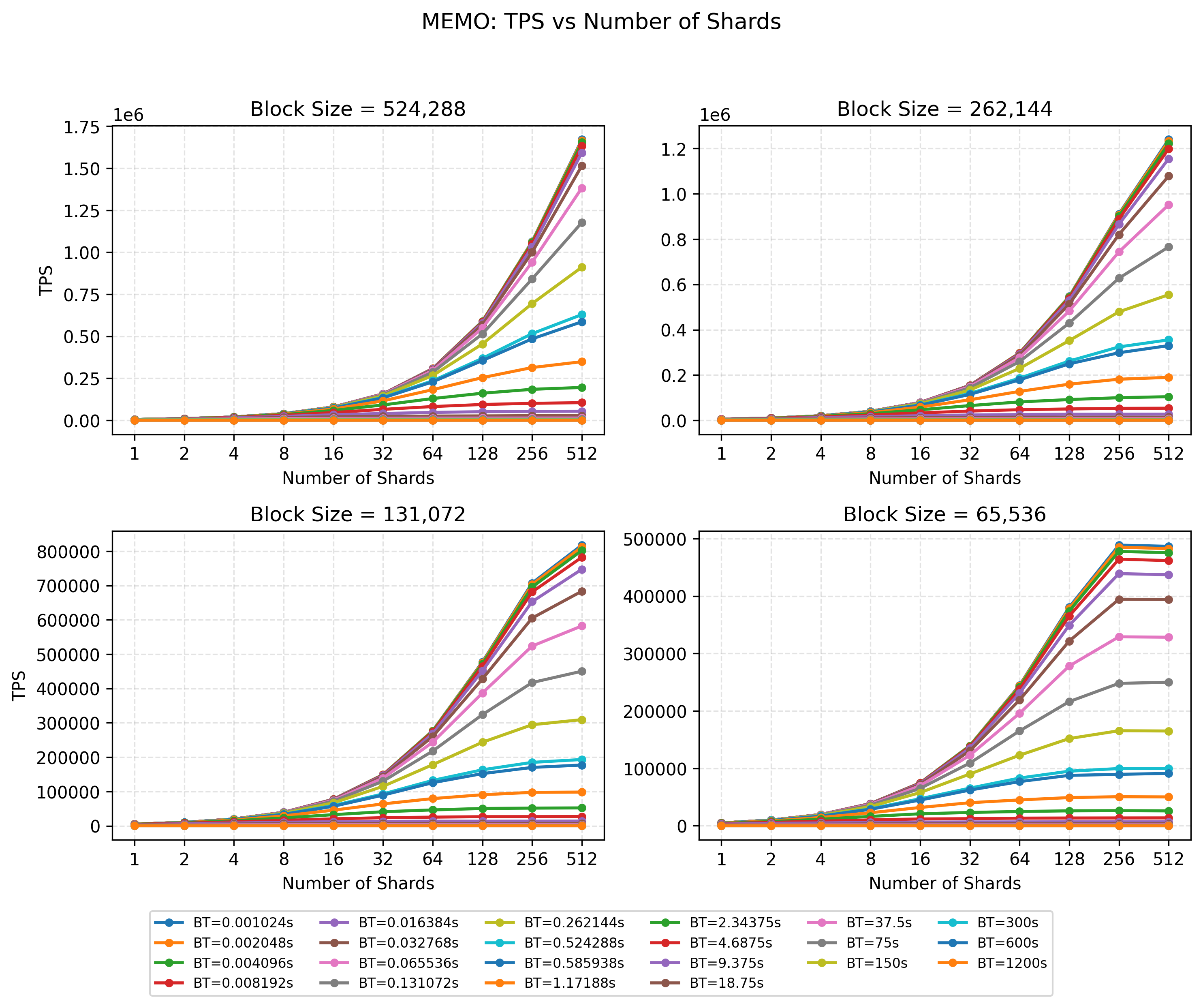}
\caption{MEMO TPS for large block sizes.}
\label{fig:tps1}
\end{figure}

\begin{figure}[!t]
\centering
\includegraphics[width=\linewidth]{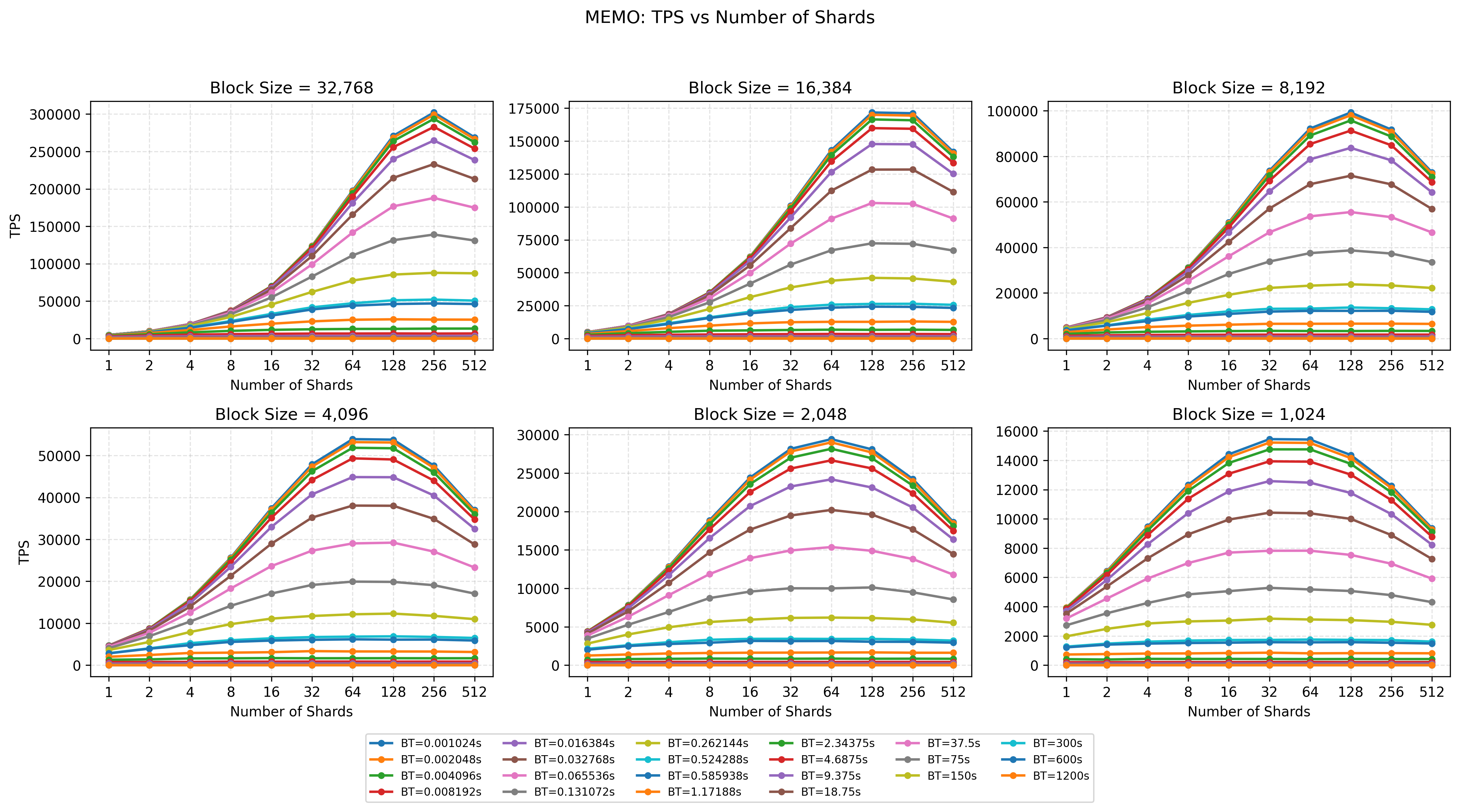}
\caption{MEMO TPS for small block sizes.}
\label{fig:tps2}
\end{figure}

Fig.~\ref{fig:tps1} and Fig.~\ref{fig:tps2} show measured TPS as a function of shard count for large and small block sizes, respectively, under local datacenter network conditions.
 
For large block sizes, throughput rises sharply with shard count at first: each additional shard contributes an independent stream of mined and verified transactions per round, so aggregate TPS scales close to linearly while the per-round coordination cost remains small relative to the volume of transactions each shard is processing. This regime continues until shard count reaches roughly 256, where throughput exceeds 1.6M TPS under local datacenter conditions, the highest measured across all configurations. Beyond this point, however, TPS growth flattens and then reverses: at 512 shards, throughput is lower than at 256 despite twice as many shards mining in parallel. The cause is the metronome's per-round challenge/proof exchange, whose message volume grows with shard count regardless of block size. As shard count increases, the leader must wait on proportionally more shard summaries before a round can be finalized, so the time spent in coordination overhead per round grows even as the time spent on useful mining and verification work per shard stays roughly constant. Past 256 shards, this coordination cost begins to consume a large enough fraction of each round that it outweighs the additional parallelism, producing a net decrease in TPS.
 
Small block sizes follow the same underlying mechanism but reach this crossover at a much lower shard count. With fewer transactions per block, each shard's mining and verification work per round is small, so the fixed per-round coordination cost represents a much larger fraction of total round time even at low shard counts. As a result, throughput peaks between 64 and 128 shards rather than 256, and the decline beyond this point is steeper: additional shards add coordination messages without adding meaningful per-shard throughput, since there are too few transactions per shard to amortize the coordination cost. This produces the diminishing, and eventually negative, returns visible at higher shard counts for small block sizes.
 
Taken together, these results show that the throughput--overhead crossover point shifts with block size: larger blocks push the optimal shard count higher because they give the metronome's fixed per-round coordination cost more useful work to amortize against, while smaller blocks saturate earlier because coordination overhead dominates sooner relative to the volume of transactions being processed. This indicates that block size and shard count cannot be tuned independently: the shard count that maximizes throughput depends directly on how much work each shard is given per round.

\section{Conclusion and Future Work}
We presented a configurable, SimPy-based discrete-event simulator for evaluating sharded blockchain architectures, modeling shard-level mining, verification, metronome-based coordination, and block dissemination under controlled network and workload assumptions. Using this simulator, we characterized throughput scaling across shard counts, block sizes, block times, and network conditions, and showed that sharding provides substantial throughput gains through parallelism, reaching 1.6M TPS at 256 shards in a local datacenter setting. These gains, however, are not unbounded: beyond a certain shard count, the metronome's per-round coordination message volume grows large enough to outweigh the benefit of additional parallelism, producing a net decrease in throughput. We further showed that this crossover point is not fixed but shifts with block size, since larger blocks give the per-round coordination cost more useful work to amortize against, allowing higher shard counts to remain productive before saturation sets in.

Future work includes calibrating the simulator against live NEAR and Ethereum testnet measurements to validate simulated TPS and finality against observed testnet behavior, extending the model to support atomic cross-shard transaction coordination and fault-tolerance scenarios, and exploring adaptive shard-count selection that adjusts to block size and network conditions at runtime. Additionally, incorporating more realistic network topologies and heterogeneous shard workloads would help assess how robust the observed throughput--overhead crossover is under conditions closer to real-world deployments.

\end{document}